\documentclass[onecolumn, aps, nofootinbib]{revtex4-2}
\usepackage{graphicx}
\usepackage{comment}
\usepackage{enumitem}
\usepackage{xcolor}
\usepackage{soul}
\usepackage{hyperref}
\hypersetup{breaklinks}
\usepackage{url}

\usepackage{multirow}
\usepackage{setspace}
\usepackage{amsmath, amssymb, amsthm}
\usepackage{cleveref}
\usepackage{physics}
\usepackage{wrapfig}

\color[rgb]{0,0,0} %black

\begin{document}

\title{Questioning the use of global estimates of reproduction numbers, with implications for policy}
%A note on misleading calculation of global R0 estimates
%The basic reproduction number and its average/

\author{Pratyush K. Kollepara}

\author{Joel C. Miller}

\affiliation{School of Engineering and Mathematical Sciences, La Trobe University, Melbourne, Australia}

\vspace{5cm}
\begin{abstract}
The basic reproduction number, $R_0$ is an important and widely used concept in the study of infectious diseases. We briefly review the recent trend of calculating the average of various $R_0$ estimates in systematic reviews aimed at estimating the basic reproduction number of SARS-CoV-2, and discuss the drawbacks and implications of using such averaging methods. Additionally, we argue that even a theoretically grounded approach such as next generation matrix could have practical impediments in its use. More generally, the practice of associating an infectious disease with a single value of $R_0$ is problematic, when the disease can, in fact have different reproduction numbers in various populations.
\end{abstract}

\maketitle

\section{Introduction}
%\vspace{-1mm}
The basic reproduction number, or $R_0$ is a widely used concept in epidemiology. It is defined as the expected number of infections caused by a typical infected individual, in a susceptible population. Thus, $R_0 = 1$ is a threshold that determines whether an outbreak can develop into an epidemic. When $R_0$ is greater than one, exponential growth in infections is expected with a positive probability, otherwise the outbreak dies out. All other factors remaining the same, a larger $R_0$ implies a higher probability of growth, with a larger exponential growth rate~\cite{Wallinga2006}. A greater $R_0$ would usually require stronger interventions to bring the epidemic under control. Thus, the knowledge of $R_0$ is of great importance to policy makers who are developing suppression strategies.

Previous theoretical work on the basic reproduction number has focused on defining this number in a mathematically rigorous manner and exploring its limitations~\cite{Wallinga2006, Diekmann1990, vandenDriessche2002, Delamater2019, Ridenhour2014}. Previous empirical work include studies that have estimated the basic reproduction number of infectious diseases in different geographical regions, using a wide range of methods. Systematic reviews of these estimates have been invaluable in understanding the various methods of estimation, and in knowing the possible range of $R_0$ values across the world.

As part of their meta-analysis, many reviews of the reproduction number of SARS-CoV-2 have calculated an ``average'' $R_0$ of the disease across multiple populations~\cite{Billah2020, Ahammed2021, Alimohamadi2020, Liu2020SARS, Liu2021Delta}. Such attempts at calculating an average reproduction number seem to implicitly assume that there exists a global basic reproduction number, and that the variation in estimates across regions is a statistical artefact or a measurement error. In this article, we critically analyze this assumption and the resulting methodology used by systematic reviews. We discuss other methods such as the next generation matrix and using the aggregated case counts, may provide a more theoretically rigorous global $R_0$. We also explore the limitations of these methods caused by delays in the spread of pandemics, finite size of the populations and the variation in generation intervals. 

\section{Mis-estimation of risk from average estimates of $R_0$}
The reproduction number depends on the biological characteristics of the pathogen, the social contact structure of the population and potentially environmental conditions. Barring the evolutionary variants of a pathogen, the biological component of $R_0$ would generally not vary across populations or regions. The social contact structure of a region is dependent on its population density and the cultural aspects of social interactions, and would affect the basic reproduction number. Therefore, if estimates of $R_0$ vary across regions, we believe this represents genuine differences between the regions rather than stochastic variation in measurements that should be averaged together with other regions. 

Regions with different values of $R_0$ require different interventions to achieve the same outcome.  This should influence the intervention choice.  Additionally, the variation limits the transferability of observations between different regions. If it is observed that an intervention reduces infections by, say, 50\% in one region, the same intervention might completely eliminate infection in one region with lower $R_0$ or have almost no impact in a region with high $R_0$.  Thus, the notion of assigning the basic reproduction number to an infectious disease at a global level can lead to poor intervention design. Averaging $R_0$ across various regions, as some systematic reviews have done \cite{Billah2020, Ahammed2021, Alimohamadi2020, Liu2020SARS, Liu2021Delta}, would mis-characterize the risks associated with an epidemic.

\begin{figure}[h]
    \centering
    \includegraphics[scale=0.95, trim={0 0 0 0}]{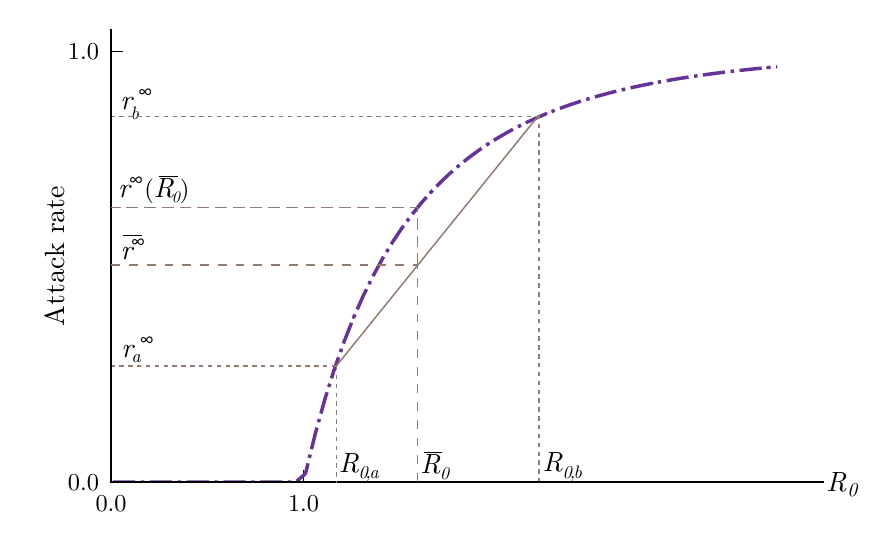}
    \caption{For $R_0>1$, the relationship between the attack rate $r^{\infty}$ and $R_0$ is convex. The points $R_{0, a}$ and $R_{0, b}$ denote the reproduction numbers in two regions $a$ and $b$, and $\overline{R_0}$ is the average of the two reproduction numbers (the average could be a weighted mean, unweighted mean or any other kind of average), and therefore lies between $R_{0, a}$ and $R_{0, b}$. The attack rate calculated using $\overline{R_0}$ is denoted by $r^{\infty}(\overline{R_0})$. The true attack rate, $\overline{r^{\infty}}$ is given by the mean of the attack rates in regions $a$ and $b$, weighted by their populations. Thus, the attack rate from the average reproduction number necessarily overestimates the attack rate. This example illustrates the implications of a convex relationship for only two regions, but it can be generalized to any number of regions \cite{Jensen1906, JensenWolfram}.}
    \label{fig:Jensen}
\end{figure}

The nonlinear relationship between final attack rate and $R_0$ of the respective local regions can lead to over-estimation of global attack rate if an average of the $R_0$ estimates is used to estimate the global attack rate (See FIG. \ref{fig:Jensen}). In addition, we illustrate two scenarios which demonstrate how average of estimates from different regions can misinform policy at local scales: 
\begin{itemize}[label=$\circ$]
    \item \textit{Averaging during a developing pandemic:} Consider a hypothetical situation where a set of four countries e.g. Australia, New Zealand, Papua New Guinea and Indonesia, differ only in terms of population density. An epidemic breaks out in Australia (which is a thinly populated region), and spreads to the thinly populated countries of New Zealand and Papua New Guinea, but also to the densely populated country of Indonesia. The reproduction number in Indonesia will be higher than the rest of the three regions, due to its high population density. If countries whose demographics are similar to Indonesia, rely on the average of the estimates to design their prevention policies, their interventions will fail to prevent adequate transmission. 
    
    \item \textit{Using averages to compare strains of pathogen:} Averaging $R_0$ values may also mis-characterize the risk of new variants of an already prevalent pathogen. A recent systematic review has compared the reproduction number of the Delta variant of SARS-CoV-2 with that of the original strain. The basic reproduction number of the Delta variant was determined using the average of estimates from China and the UK, and was compared with the average of reproduction number of the original strain from China \cite{Liu2021Delta}. Since the social contact structure and population behaviour differ in China and the UK, the appropriate way to compare reproduction numbers of different variants is by using estimates from a single region. Furthermore, if a new strain has an increased reproduction number due to a variation in the biology of the pathogen, this increase in $R_0$ will not necessarily be reflected in the average $R_0$ sampled from different regions if the sample is dominated by regions with sparse social contact structures.
\end{itemize}
 
\section{Other methods to estimate global $R_0$}
\begin{figure*}[t]
    \centering
    \includegraphics[scale=1, trim={0 0 0 0}]{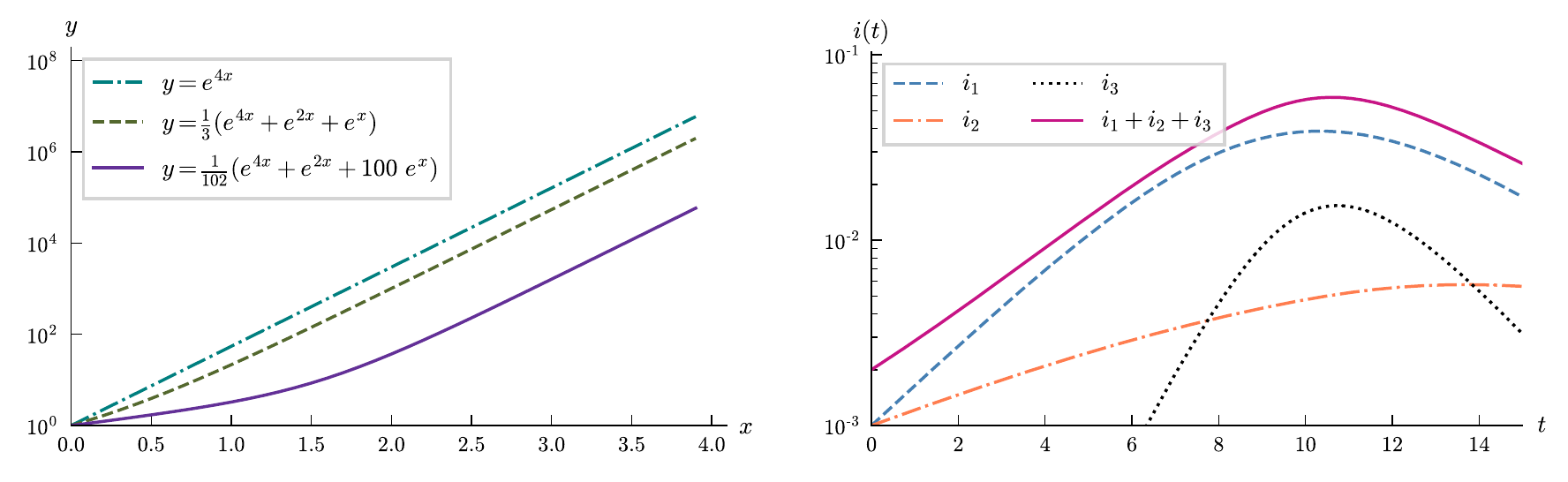}
    \caption{Left: The growth rate of the sum of exponential functions converges to the largest exponent. Right: The fastest exponential growth is not reflected in the aggregated case counts if the sub-populations are weakly coupled, and have finite size. The third group, labelled by $i_3$ has the fastest growth, yet its exponent is not reflected in the aggregate because the epidemic had a delayed start and because the sub-population had the smallest size.}
    \label{fig:exp}
\end{figure*}

The problem of finding the global $R_0$ of an infectious disease is a specific case of the more general problem of finding the $R_0$ of a collection of populations, each with its own internal $R_0$. A commonly used method to estimate $R_0$ at national or sub-national level is to use the time series of aggregated case counts. When the size of populations is sufficiently large, the population with the fastest exponential growth will dominate the time series eventually (See FIG. \ref{fig:exp}). This is predicted by the next generation matrix \cite{Diekmann1990, vandenDriessche2002}, from which we find that the global $R_0$ of a collection of weakly-coupled sub-populations (transmission between regions is negligible compared to transmission within regions) is the highest observed reproduction number among the regions. While this approach recognizes the variation in reproduction numbers of the sub-populations, there are practical limitations to it.

In an emerging pandemic, epidemics in different regions would not start at the same time. For example, travelers may introduce infection into a more affluent community, and we may not observe growth in a poorer subpopulation until much later.  Therefore, an estimate consistent with the next generation matrix approach, will not be determined until epidemics have established in a large number of regions. Furthermore, since the size of populations is finite, it is possible that the largest exponential growth rate will not be observable in the aggregated data (See FIG. \ref{fig:exp} and simulation details \footnote{The figure is generated from an SIR model with three communities. The equation for active infections in the $k^{\text{th}}$ is given by $\dv{i_k}{t} = s_k(\beta_k i_k + c\Sigma_{l \neq k} \beta_l i_l) - \gamma i_k$. The sizes of the three communities are $(n_1, n_2, n_3) = (0.6, 0.3, 0.1)$, model parameters are $(\beta_1, \beta_2, \beta_3, \gamma, c) = (2.5, 4, 20, 1, 10^{-3})$, and initial conditions are $r_k = 0$, $(s_1, s_2, s_3) = (10^{-3}, 10^{-3}, 10^{-6})$. The parameter $c$ is the coupling between communities.}). Thus, policies designed on the basis of early behaviour of epidemics (including the $R_0$), such as vaccination targets and time-lines may fall short if the delayed epidemics are faster than the early average of estimates.

\begin{figure}[h]
    \centering
    \includegraphics[scale=0.69, trim={0 0 0 0}]{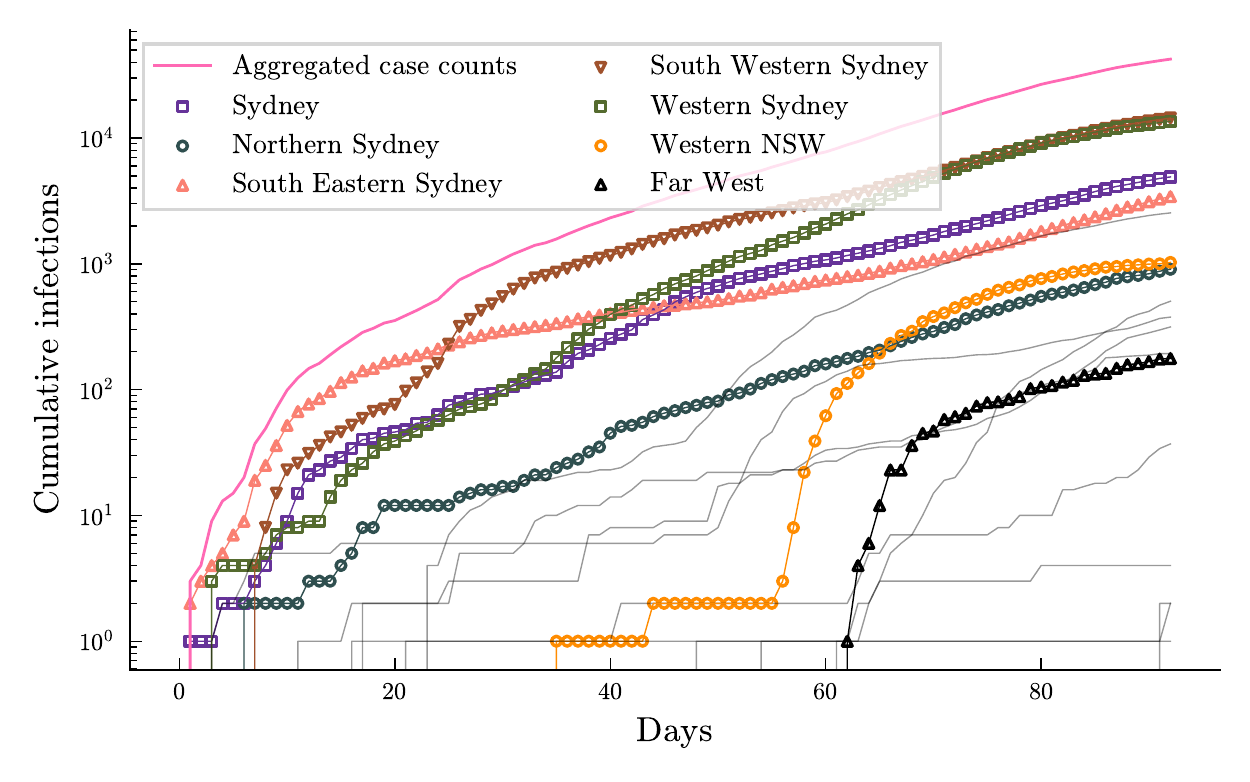}
    \caption{Cumulative time series of infections in the local health districts of New South Wales, Australia, during the second wave. Within the city of Sydney, we observe different rates of growth. Regions outside Sydney such as Western NSW and Far West exhibit the highest growth rates, but their epidemic was delayed and due to their finite populations, the exponential growth did not get reflected in the aggregated case counts. The grey line-plots are from the rest of the local health districts in NSW. Day 0 refers to 2021-06-15. Data obtained from \cite{NSW_data}}
    \label{fig:NSW}
\end{figure}

We demonstrate these limitations using time series of cumulative infections from various health districts of New South Wales (NSW) in Australia. Figure \ref{fig:NSW} shows that the growth rates can be different even inside Sydney. Districts outside Sydney, such as Western NSW and Far West had a delayed epidemic, and had significantly faster growth than the epidemics in the rest of the state. The delay in the epidemic, compounded by the finite size of the populations (and a smaller population in case of Far West), led to the aggregated case counts not reflecting these higher exponential growth rates.

Finally, even if all the populations are sufficiently large and the epidemics start at the same time, the exponent observed in the aggregated case counts may not correspond to the largest $R_0$ if the generation intervals are not identical across the sub-populations. The variation in the generation interval distribution can also be attributed to to social contact structure. For example, populations with larger household sizes might have shorter generation intervals. Thus, it is not necessary that the $R_0$ calculated from aggregated case numbers is consistent with the next generation matrix.

\section{Conclusion}
In closing, we re-iterate that the variation in estimates of the reproduction number from different regions of the world is not a statistical error, but the basic reproduction number or $R_0$ depends on the social contact structure and environmental conditions. Therefore, we recommend that systematic reviews should continue to explore the range of $R_0$ values and the various methods used to estimate it. But finding an ``$R_0$ of the disease" or assigning a single value of $R_0$ to a disease should not be the focus of reviews. Indeed, a systematic review of measles also highlighted the need for countries to use local estimates over averages \cite{Guerra2017}. 

The next generation matrix approach may provide a theoretically sound value of the global $R_0$ of the disease, but there are several practical limitations in its estimation, and such a global estimate would nonetheless be of limited use. 

\section{Acknowledgements}
We thank Dr.{} Rebecca Chisholm for comments and discussion.

\bibliography{references}{}
\bibliographystyle{naturemag}

\end{document}